\documentclass[12pt,letterpaper]{article}

\usepackage{amsmath}
\usepackage{amsfonts}
\usepackage{amssymb}

\allowdisplaybreaks

\newtheorem{lem}{Lemma}

\newtheorem{defin}{Definition}
\newtheorem{prop}{Proposition}
\newtheorem{cor}{Corollary}
\newtheorem{rem}{Remark}

\newtheorem{ppty}{Property}

\begin{document}

\title{\textsc{Preference-Based Unawareness}}

\author{Burkhard C. Schipper\thanks{Department of Economics, University of California, Davis. Email: bcschipper@ucdavis.edu \newline I am very grateful to an anonymous referee for extremely helpful comments. This paper is closely related to prior work on unawareness with my friends, Aviad Heifetz and Martin Meier. All mistakes are mine. Financial support through NSF SES-0647811 is gratefully acknowledged.}}

\date{This Version: February 18, 2013 \\ First Version: April 7, 2011}

\maketitle

\begin{abstract} Morris (1996, 1997) introduced preference-based definitions of knowledge and belief in standard state-space structures. This paper extends this preference-based approach to unawareness structures (Heifetz, Meier, and Schipper, 2006, 2008). By defining unawareness and knowledge in terms of preferences over acts in unawareness structures and showing their equivalence to the epistemic notions of unawareness and knowledge, we try to build a bridge between decision theory and epistemic logic. Unawareness of an event is characterized behaviorally as the event being null and its negation being null.
\newline
\textbf{JEL-Classifications:} C70, C72, D80, D82. \bigskip
\end{abstract}







\newpage

\section{Introduction}

Unawareness refers to the lack of conception rather than the lack of information. There is a fundamental difference between not knowing about which events obtain and the inability to conceive of some events. Unawareness is an interdisciplinary topic that fascinates economists, computer scientists, logicians, and philosophers alike. Traditionally, computer scientists, logicians and philosophers are interested in epistemic models while most economists are mainly interested in the behavioral implications. In the literature, unawareness has been defined epistemically using syntactic and semantic approaches.\footnote{For a bibliography see http://www.econ.ucdavis.edu/faculty/schipper/unaw.htm} While epistemic characterizations are conceptually insightful, the behavioral content of unawareness is less clear. Can unawareness be characterized behaviorally?

A first attempt to answer this question is presented in Schipper (2012). There we apply the lattice state-space structure of Heifetz, Meier, and Schipper (2006, 2008, 2013) to the Anscombe-Aumann approach to subjective utility theory and characterize awareness-dependent subjective expected utility. This framework allows us to distinguish unawareness of an event from the event being Savage null and thus we are able to show behavioral implications of unawareness. Yet, our approach in Schipper (2012) has many shortcomings. For instance, as in almost any decision theoretic framework, we consider \emph{states of nature} rather than \emph{states of the world}. That is, states describe just lists of physical events but are silent on the state of mind of the decision maker. While such an approach is sensible in a purely decision theoretic framework, where bets are placed on physical events rather than (higher order) beliefs of some players, it does not allow for characterizing behaviorally - at least in principle - the intricate epistemic properties of knowledge and awareness across the lattice structure introduced in Heifetz, Meier, and Schipper (2006, 2008). For instance, introspection of knowledge requires bets on events in which the decision maker knows an event.

Morris (1996, 1997) showed for standard state-spaces with states of the world that properties of knowledge and belief can be - at least in principle - characterized behaviorally. In this paper, we extend his approach to unawareness structures and show how to characterize behaviorally properties of the possibility correspondence introduced in Heifetz, Meier, and Schipper (2006). Moreover, we show how to characterize preference-based operators of knowledge, awareness, and unawareness.

This work is directly related to a growing literature on unawareness. Heifetz, Meier, and Schipper (2006, 2008, 2013) introduced a syntax-free semantics of unawareness using state-spaces familiar to economists, decision theorists, and game theorists. Apart from having a syntax-free semantics, Heifetz, Meier, and Schipper (2006, 2008) generalize Modica and Rustichini (1999) to the multi-agent case. The precise connection between Heifetz, Meier, and Schipper (2006) on one hand and Modica and Rustichini (1999) and earlier work in computer science by Fagin and Halpern (1988) on the other hand is understood from Halpern (2001), Halpern and R\^{e}go (2008, 2012), and Heifetz, Meier, and Schipper (2008). Galanis (2012) relaxes some properties of the possibility correspondence in Heifetz, Meier, and Schipper (2006). The connection between Heifetz, Meier, and Schipper (2006, 2008) and Galanis (2012) is explored in Galanis (2011). Li (2009) presents an alternative to modeling unawareness in the single-agent case. Heinsalu (2012) explores the connection between Li (2009) and the rest of the literature. Finally, Feinberg (2012) presents an approach of modeling unawareness in the context of games. The precise connection between Feinberg (2012) and the rest of the literature is yet to be explored. The literature addressed also awareness of unawareness, see Board and Chung (2011), Halpern and R\^{e}go (2009, 2012a), Sillari (2008), {\AA}gotnes and Alechina (2007), and Walker (2013). The relationship between Board and Chung (2011) and Heifetz, Meier, and Schipper (2006) is studied in Board, Chung, and Schipper (2011). The relationship between Halpern and Rego (2009, 2012a) and Heifetz, Meier and Schipper (2006) is explored in Halpern and Rego (2012a). See Section 4 for further discussions of the related literature.

In the next section we provide a brief exposition of unawareness structures and review properties of knowledge and awareness. In Section 3 we introduce the decision theoretic framework and provide the results. We conclude with a discussion in Section 4. Proofs are collected in the appendix.

\section{Unawareness Structures}

\subsection{State-Spaces}

Let $\mathcal{S}=\left\{ S_{\alpha }\right\} _{\alpha \in \mathcal{A}}$ be a
finite lattice of disjoint \emph{state-spaces}, with the partial order $
\preceq $ on $\mathcal{S}$. For simplicity we assume in this paper that each $S$ is finite. If $S_{\alpha }$ and $S_{\beta }$ are such that $S_{\alpha }\succeq S_{\beta }$ we say that \textquotedblleft $S_{\alpha }$ is
more expressive than $S_{\beta }$ -- states of $S_{\alpha }$ describe
situations with a richer vocabulary than states of $S_{\beta }$
\textquotedblright .\footnote{Here and in what follows, phrases within
quotation marks hint at intended interpretations, but we emphasize that these
interpretations are not part of the definition of the set-theoretic structure.}
Denote by $\Omega =\bigcup_{_{\alpha \in \mathcal{A}}}S_{\alpha }$ the disjoint union of
these spaces.

Spaces in the lattice can be more or less ``rich'' in terms of facts that may
or may not obtain in them. The partial order relates to the ``richness'' of
spaces. The upmost space of the lattice may be interpreted as the ``objective''
state-space. Its states encompass full descriptions.

\subsection{Projections}

For every $S$ and $S^{\prime }$ such that $S^{\prime }\succeq S$, there is a surjective projection $r_{S}^{S^{\prime }}:S^{\prime }\rightarrow
S$, where $r_S^S$ is the identity. (``$r_{S}^{S^{\prime }}\left( \omega
\right) $ is the restriction of the description $\omega $ to the more
limited vocabulary of $S$.'') Note that the cardinality of $S$ is smaller
than or equal to the cardinality of $S^{\prime }$. We require the
projections to commute: If $S^{\prime \prime }\succeq S^{\prime }\succeq S$
then $r_{S}^{S^{\prime \prime }}=r_{S}^{S^{\prime }}\circ
r_{S^{\prime}}^{S^{\prime \prime }}$. If $\omega \in S^{\prime }$, denote $%
\omega_{S}=r_{S}^{S^{\prime }}\left( \omega \right) $. If $D \subseteq
S^{\prime }$, denote $D_{S}=\left\{ \omega_{S}: \omega \in D \right\}$.

Projections \textquotedblleft translate\textquotedblright\ states in
\textquotedblleft more expressive\textquotedblright\ spaces to states in
\textquotedblleft less expressive\textquotedblright\ spaces by
\textquotedblleft erasing\textquotedblright\ facts that can not be expressed
in a lower space.

These surjective projections may embody Savage's idea that ``(i)t may be well, however, to emphasize that a state of the smaller world corresponds not to a state of the larger, but to a set of states'' (Savage, 1954, p. 9).

\subsection{Events}

Denote $g(S)=\left\{ S^{\prime }:S^{\prime }\succeq S\right\}$. For $D \subseteq S$, denote $D^{\uparrow }=\bigcup_{S^{\prime }\in g(S)}\left(r_{S}^{S^{\prime }}\right) ^{-1}\left( D\right)$. (``All the extensions of descriptions in $D$ to at least as expressive vocabularies.'')

An \textit{event} is a pair $\left( E, S\right) $, where $E = D^{\uparrow }$ with $D \subseteq S$, where $S \in \mathcal{S}$. $D$ is called the \textit{base} and $S$ the \textit{base-space} of $\left( E, S\right) $, denoted by $S(E)$. If $E\neq \emptyset $, then $S$ is uniquely determined by $E$ and, abusing notation, we write $E$ for $\left( E, S\right) $. Otherwise, we write $\emptyset^{S}$ for $\left( \emptyset ,S\right) $. Note that not every subset of $\Omega$ is an event.

Some fact may obtain in a subset of a space. Then this fact should be also
\textquotedblleft expressible\textquotedblright\ in \textquotedblleft more
expressive\textquotedblright\ spaces. Therefore the event contains not only
the particular subset but also its inverse images in \textquotedblleft more
expressive\textquotedblright\ spaces.

Let $\Sigma $ be the set of events of $\Omega $. Note that unless $\mathcal{S}$ is a singleton, $\Sigma$ is not an algebra on $\Omega$ because it contains distinct vacuous events $\emptyset^S$ for all $S \in \mathcal{S}$. These vacuous events correspond to contradictions with differing ``expressive power''.

\subsection{Negation}\label{negation}

If $(D^{\uparrow },S)$ is an event where $D\subseteq S$, the negation $\lnot
(D^{\uparrow },S)$ of $(D^{\uparrow },S)$ is defined by $\lnot (D^{\uparrow
},S):=((S\setminus D)^{\uparrow },S)$. Note, that by this definition, the
negation of a (measurable) event is a (measurable) event. Abusing notation,
we write $\lnot D^{\uparrow }:=(S\setminus D)^{\uparrow }$. Note that by our
notational convention, we have $\lnot S^{\uparrow }=\emptyset ^{S}$ and $\lnot \emptyset ^{S}=S^{\uparrow },$ for each space $S\in \mathcal{S}$. The
event $\emptyset ^{S}$ should be interpreted as a \textquotedblleft logical
contradiction phrased with the expressive power available in $S$.\textquotedblright\ $\lnot D^{\uparrow }$ is typically a proper subset of
the complement $\Omega \setminus D^{^{\uparrow }}$. That is, $\left(
S\setminus D\right) ^{\uparrow }\subsetneqq \Omega \setminus D^{^{\uparrow
}} $.

Intuitively, there may be states in which the description of an event
$D^{\uparrow }$ is both expressible and true -- these are the states in
$D^{\uparrow }$; there may be states in which its description is expressible
but false -- these are the states in $\lnot D^{\uparrow }$; and there may be
states in which neither its description nor its negation are expressible --
these are the states in
\begin{equation*}
\Omega \setminus \left( D^{\uparrow }\cup \lnot D^{\uparrow }\right) =\Omega
\setminus S\left( D^{\uparrow} \right)^{\uparrow }.
\end{equation*}
Thus our structure is not a standard state-space model in the sense of
Dekel, Lipman, and Rustichini (1998).

\subsection{Conjunction and Disjunction}

If $\left\{ \left( D_{\lambda }^{\uparrow }, S_{\lambda }\right) \right\}_{\lambda \in L}$ is a collection of events (with $D_{\lambda }\subseteq S_{\lambda }$, for $\lambda \in L$), their conjunction $\bigwedge_{\lambda \in L}\left(D_{\lambda }^{\uparrow }, S_{\lambda }\right) $ is defined by $\bigwedge_{\lambda \in L}\left(D_{\lambda }^{\uparrow }, S_{\lambda }\right) :=\left( \left(\bigcap_{\lambda \in L} D_{\lambda }^{\uparrow }\right) ,\sup_{\lambda \in L}S_{\lambda }\right) $. Note, that since $\mathcal{S}$ is a \textsl{complete} lattice, $\sup_{\lambda \in L} S_{\lambda}$ exists. If $S = \sup_{\lambda \in L} S_{\lambda }$, then we have $\left( \bigcap_{\lambda \in L} D_{\lambda }^{\uparrow }\right) =\left( \bigcap_{\lambda \in L}\left( \left( r_{S_{\lambda }}^{S}\right) ^{-1}\left( D_{\lambda }\right) \right) \right)^{\uparrow }$. Again, abusing notation, we write $\bigwedge_{\lambda \in L} D_{\lambda }^{\uparrow } := \bigcap_{\lambda \in L} D_{\lambda }^{\uparrow }$ (we will therefore use the conjunction symbol $\wedge $ and the intersection symbol $\cap$ interchangeably).

We define the relation $\subseteq $ between events $\left( E,S\right) $ and $
\left( F, S^{\prime }\right) ,$ by $\left( E,S\right) \subseteq \left( F,
S^{\prime }\right) $ if and only if $E\subseteq F$ as sets \emph{and }$
S^{\prime }\preceq S.$ If $E\neq \emptyset $, we have that $\left(
E,S\right) \subseteq \left( F,S^{\prime }\right) $ if and only if $
E\subseteq F$ as sets. Note however that for $E=\emptyset ^{S}$ we have $
\left( E,S\right) \subseteq \left( F,S^{\prime }\right) $ if and only if $
S^{\prime }\preceq S.$ Hence we can write $E\subseteq F$ instead of $\left(
E, S\right) \subseteq \left( F, S^{\prime }\right) $ as long as we keep in
mind that in the case of $E=\emptyset ^{S}$ we have $\emptyset ^{S}\subseteq
F$ if and only if $S\succeq S(F)$. It follows from these definitions that
for events $E$ and $F$, $E\subseteq F$ is equivalent to $\lnot F\subseteq
\lnot E$ only when $E$ and $F$ have the same base, i.e., $S(E)=S(F)$.

The disjunction of $\left\{ D_{\lambda }^{\uparrow }\right\} _{\lambda \in
L} $ is defined by the de Morgan law $\bigvee_{\lambda \in L}D_{\lambda
}^{\uparrow }=\lnot \left( \bigwedge_{\lambda \in L}\lnot \left( D_{\lambda
}^{\uparrow }\right) \right) $. Typically $\bigvee_{\lambda \in L}D_{\lambda
}^{\uparrow }\subsetneqq \bigcup_{\lambda \in L}D_{\lambda }^{\uparrow }$,
and if all $D_{\lambda }$ are nonempty we have that $\bigvee_{\lambda \in
L}D_{\lambda }^{\uparrow }=\bigcup_{\lambda \in L}D_{\lambda }^{\uparrow }$
holds if and only if all the $D_{\lambda }^{\uparrow }$ have the same
base-space. Note, that by these definitions, the conjunction and disjunction of events is an event.

\subsection{Possibility Correspondence}

The set of states that a decision maker considers possible at a state is modeled by a \emph{possibility correspondence} $\Pi: \Omega \longrightarrow 2^{\Omega}$ that satisfies the following properties:

\begin{enumerate}
\item[0.] \emph{Confinement:} If $\omega \in S$ then $\Pi(\omega)\subseteq S'$ for some $S' \preceq S$.

\item[1.] \emph{Generalized Reflexivity:} $\omega \in \Pi^{\uparrow}(\omega)$ for every $\omega \in \Omega$.\footnote{Here and in what follows, we abuse notation slightly and write $\Pi^{\uparrow}(\omega)$ for $\left(\Pi(\omega )\right)^{\uparrow}$.}

\item[2.] \emph{Stationarity:} $\omega^{\prime }\in \Pi\left( \omega \right)$ implies $\Pi\left( \omega^{\prime }\right) = \Pi\left( \omega\right)$.

\item[3.] \emph{Projections Preserve Awareness:} If $\omega \in S'$, $\omega \in \Pi(\omega)$ and $S \preceq S'$ then $\omega_{S} \in \Pi(\omega_{S})$.

\item[4.] \emph{Projections Preserve Ignorance:} If $\omega \in S'$ and $S \preceq S'$ then $\Pi^{\uparrow }(\omega) \subseteq \Pi^{\uparrow}(\omega_{S})$.

\item[5.] \emph{Projections Preserve Knowledge:} If $S \preceq S' \preceq S^{''}$, $\omega \in S^{''}$ and $\Pi(\omega)\subseteq S'$ then $\left( \Pi( \omega) \right)_{S}= \Pi( \omega_{S})$.
\end{enumerate}

For 5., we could have assumed $\supseteq $ and deduce $=$ from $\supseteq $,
3., and the other properties.

\begin{rem} Generalized Reflexivity implies that if $S' \preceq S$, $\omega \in S$ and $\Pi(\omega) \subseteq S'$, then $r_{S'}^{S}(\omega) \in \Pi(\omega)$. In particular, we have $\Pi(\omega) \neq \emptyset$, for all $\omega \in \Omega$.
\end{rem}

\begin{rem} Projections Preserve Ignorance and Confinement imply that if $S' \preceq S$, $\omega \in S$ and $\Pi(\omega_{S'}) \subseteq S^{''}$, then $\Pi(\omega) \subseteq S^*$ for some $S^*$ with $S^{''} \preceq S^*$.
\end{rem}

\begin{rem}\label{imply3} Projections Preserve Knowledge and Confinement imply Property $3$.
\end{rem}

Generalized Reflexivity and Stationarity are the analogues of the
partitional properties of the possibility correspondence in partitional
information structures. In particular, Generalized Reflexivity will yield
the truth property (that what an individual knows indeed obtains -- property
(iii) in Proposition~\ref{Kproperties}); Stationarity will guarantee the
introspection properties (that an individual knows what she knows --
property (iv) in Proposition~\ref{Kproperties}, and that an individual knows
what she ignores provided she is aware of it -- Property 5. in Proposition~\ref{AK}).

Properties 3. to 5. of the possibility correspondence guarantee the coherence of the knowledge and the awareness of individuals down the lattice structure. They compare the possibility sets of an individual in a state $\omega$ and its projection $\omega_{S}$. The properties guarantee that, first, at the projected state $\omega_{S}$ the individual knows nothing she does not know at $\omega$, and second, at the projected state $\omega_{S}$ the individual is not aware of anything she is unaware of at $\omega$ (Projections Preserve Ignorance). Third, at the projected state $\omega_{S}$ the individual knows every event she knows at $\omega$, provided that this event is based in a space lower than or equal to $S$ (Projections Preserve Knowledge). Fourth, at the projected state $\omega_{S}$ the individual is aware of every event she is aware of at $\omega$, provided that this event is based in a space lower than or equal to $S$ (Projections Preserve Awareness).

\subsection{Knowledge}

\begin{defin}\label{K} The decision maker's \emph{knowledge operator} on events $E$ is defined, as usual, by
\begin{equation*}
K(E):=\left\{ \omega \in \Omega :\Pi(\omega) \subseteq E \right\},
\end{equation*}
if there is a state $\omega$ such that $\Pi(\omega) \subseteq E$, and
by
\begin{equation*}
K(E):=\emptyset^{S(E)}
\end{equation*}
otherwise.
\end{defin}

The following two propositions are proved in Heifetz, Meier, and Schipper
(2006):

\begin{prop}\label{Kbase} If $E$ is an event, then $K(E)$ is an $S(E)$-based event.
\end{prop}

\begin{prop}\label{Kproperties} The Knowledge operator $K$ has the following
properties:

\begin{itemize}
\item[(i)] Necessitation: $K(\Omega) = \Omega$,

\item[(ii)] Conjunction: $K\left( \bigcap_{\lambda\in L} E_{\lambda
}\right) = \bigcap_{\lambda\in L} K\left( E_{\lambda} \right)$,

\item[(iii)] Truth: $K(E)\subseteq E$,

\item[(iv)] Positive Introspection: $K(E)\subseteq KK(E)$,

\item[(v)] Monotonicity: $E \subseteq F$ implies $K(E)\subseteq K(F)$,

\item[(vi)] $\lnot K(E)\cap\lnot K\lnot K(E)\subseteq\lnot
K\lnot K\lnot K(E)$.
\end{itemize}
\end{prop}

Proposition~\ref{Kproperties} says that the knowledge operator has all the
strong properties of knowledge in partitional information structures, except
for the weakening (vi) of the negative introspection property. Negative
introspection -- the property $\lnot K(E)\subseteq K\lnot K(E)$
that when an individual does not know an event, she knows she does not know
it -- obtains only when the individual is also aware of the event (see
Property 5. of the next proposition).

\subsection{Awareness and Unawareness}

\begin{defin}\label{A} The decision maker's \emph{awareness operator} from events to events is defined by
\begin{equation*}
A(E) = \{\omega \in \Omega : \Pi(\omega) \subseteq
S(E)^{\uparrow} \}
\end{equation*}
if there is a state $\omega$ such that $\Pi(\omega) \subseteq
S(E)^{\uparrow}$, and by
\begin{equation*}
A(E) = \emptyset^{S(E)}
\end{equation*}
otherwise. The \emph{unawareness operator} is then naturally defined by
\begin{equation*}
U(E) = \lnot A(E).
\end{equation*}
\end{defin}

Alternatively, we could follow Modica and Rustichini (1999) in defining the
\emph{unawareness operator} by
\begin{equation*}
U(E) = \lnot K(E)\cap \lnot K\lnot K(E).
\end{equation*} Heifetz, Meier, and Schipper (2008, Remark 6) show that the two definitions are indeed equivalent. Note that by Proposition~\ref{Kbase} and the definition of the negation, we have
\begin{equation*}
A(E) = K(E)\cup K \lnot K(E).
\end{equation*}

The following proposition is proved in Heifetz, Meier, and Schipper (2006):

\begin{prop}
\label{AK} The following properties of knowledge and awareness obtain:

\begin{enumerate}
\item $KU$ Introspection: $KU(E)=\emptyset ^{S(E)}$,

\item $AU$ Introspection: $U(E)=UU(E)$,

\item Weak Necessitation: $A(E)=K\left( S\left( E\right) ^{\uparrow
}\right) $,

\item Strong Plausibility: $U(E)=\bigcap_{n=1}^{\infty }\left( \lnot
K\right) ^{n}(E)$,

\item Weak Negative Introspection: $\lnot K(E)\cap A\lnot K(E) =
K\lnot K(E)$,

\item Symmetric: $A(E) = A(\lnot E)$,

\item $A$-Conjunction: $\bigcap_{\lambda \in L}A\left( E_{\lambda
}\right) =A\left( \bigcap_{\lambda \in L}E_{\lambda }\right) $,

\item $AK$-Self Reflection: $A(E) = AK(E)$,

\item $AA$-Self Reflection: $A(E) = AA(E)$,

\item $A$-Introspection: $A(E) = KA(E)$.
\end{enumerate}
\end{prop}

Properties 1. to 4. have been proposed by Dekel, Lipman and Rustichini
(1998), properties 6. to 9. by Modica and Rustichini (1999), and properties
5. to 9. by Fagin and Halpern (1988) and Halpern (2001). $A$-Introspection
is the property that an individual is aware of an event if and only if she
knows she is aware of it.

\section{Preference-Based Knowledge and Awareness}

So far, we just outlined unawareness structures introduced in Heifetz, Meier, and Schipper (2006, 2008). In this section, we add decision theoretic primitives and characterize the possibility correspondence, knowledge, awareness and unawareness by choices. As in Morris (1996, 1997), from now on we restrict ourselves to finite $\Omega$ (see also Fn.~\ref{finite}). This assumption may be defended by the fact that most choice experiments take place in a finite context.

\subsection{Acts}

An act is a function $f: \Omega \longrightarrow \mathbb{R}$, where $f(\omega) \in \mathbb{R}$ can be thought of the money prize at state $\omega$.

Note that different from Savage acts, $f$ is not defined on just one state-space but on the union of spaces $\Omega$. This is interpreted as follows: Let's say an individual considers investing in a firm (e.g., the act $f$). A firm can be viewed as a bundle of uncertain opportunities and liabilities. The decision maker may perceive only a subset of opportunities and liabilities depending on her awareness level that may be influenced by her prior experience or her reading of the ``fine prints'' of the ``share sales and purchase agreement''. In our setting, the  opportunities and liabilities are represented by money prizes contingent on states, i.e., acts. Some of the opportunities and liabilities (i.e., events) may not be expressible in some of the spaces in $\mathcal{S}$. That is, they are not perceived when having certain awareness levels. So an act denotes simultaneously more or less rich descriptions of those opportunities and liabilities. It is essentially a label of the action ``buy the firm''. Which opportunities and liabilities are perceived by the decision maker, i.e., the awareness level of the decision maker, will be captured by the preferences over acts introduced below.

For any event $E$ and acts $f$ and $g$, define a \emph{composite act}
$f_E g$ by $$f_E g (\omega) = \left\{\begin{array}{cl} f(\omega) & \mbox{ if } \omega \in E \\ g(\omega) & \mbox{ otherwise. } \end{array} \right.$$ Note that different from composite acts in the Savage approach, $g$ is not only prescribed on the negation of $E$ but also on all states that are neither in $E$ nor in $\neg E$. Different from acts defined on a standard state-space, we have in general $f_E g \neq g_{\neg E} f$.

For any collection of pairwise disjoint events $E_1, E_2, ..., E_n \in \Sigma$ and acts $f^1, f^2, ..., f^n, g \in \mathcal{A}$, let $f^1_{E_1} f^2_{E_2} ... f^n_{E_n} g$ denote the composite act that yields $f^i(\omega)$ if $\omega \in E_i$ for $i = 1, ..., n$, and $g(\omega)$ otherwise.

Let $\mathcal{A}$ denote the set of all acts.

Note that we do not impose a measurability condition on acts in the sense that for any $f \in \mathcal{A}$ and $x \in \mathbb{R}$,  the set of states $\{\omega \in \Omega : f(\omega) = x \}$ is an event in $\Sigma$ as defined previously. While such a measurability assumption may be justified in some applications, it may not be applicable in general. The speculative trade example in Heifetz, Meier, and Schipper (2006) is one instance.\footnote{There, the set of states where the value of the firm is 90 dollars (for instance) in not an event in the sense of our event structure. Thus, the act ``buy the firm'' would not satisfy such a measurability condition. See also Heifetz, Meier, and Schipper (2013) and Meier and Schipper (2010) for a detailed analysis of speculative trade under unawareness.} A practical framework of decision making under unawareness should not rule out such examples.

In the following we will consider also composite acts of the form
\begin{eqnarray*} f_{\{\omega\}} g (\omega') = \left\{ \begin{array}{ll} f(\omega') & \mbox{ if } \omega = \omega' \\ g(\omega') & \mbox{ otherwise.} \end{array} \right.
\end{eqnarray*} Although $\{\omega\}$ may not be an event in the unawareness structure, we still have $f_{\{\omega\}} g \in \mathcal{A}$. To see this note that for every $f, g \in \mathcal{A}$ we can define an act $h \in \mathcal{A}$ such that $h(\omega) = f(\omega)$ and $h(\omega') = g(\omega')$ for $\omega' \neq \omega$. Then $f_{\{\omega\}} g = h_{\{\omega \}^{\uparrow}} g$ and clearly $h_{\{\omega \}^{\uparrow}} g \in \mathcal{A}$.

\subsection{Preferences}

We follow Morris (1996, 1997) in defining a binary relation $\succsim(\omega)$ on acts in $\mathcal{A}$ for each $\omega \in \Omega$. We assume that for any $\omega \in \Omega$, $\succsim(\omega)$ is a preference relation, i.e., $\succsim(\omega)$ is reflexive, complete, and transitive. That is, at each state the decision maker is assumed to have a preference relation. $f \succsim(\omega) \ g$ means that at state $\omega$, the decision maker prefers act $f$ over act $g$.\footnote{Note that $f \succsim(\omega) \ g$ does not necessarily imply that the decision maker would prefer $f$ over $g$ if he \emph{believes} that the state is $\omega$.} Reflexivity means that every act is as good as itself. Completeness is a strong assumption. It requires that the decision maker can rank any two acts. Transitivity is sufficient to rule out ``money pumps'', i.e., choice cycles along which the decision maker could be exploited.

As usual, strict preference, $\succ(\omega)$, is defined on $\mathcal{A}$ by $\succsim(\omega)$ and not $\precsim(\omega)$. Indifference, $\sim(\omega)$, is defined on $\mathcal{A}$ by $\succsim(\omega)$ and $\precsim(\omega)$.

\subsection{Preference-Based Possibility Correspondence}

Define a preference-based correspondence $\tilde{\Pi} : \Omega \longrightarrow 2^{\Omega}$ by
\begin{eqnarray} \tilde{\Pi}(\omega) = \{ \omega' \in \Omega : f_{\{\omega'\}} g \succ(\omega) \ h_{\{\omega'\}} g \mbox{ for some } f, g, h \in \mathcal{A} \}.
\end{eqnarray}

Intuitively, at state $\omega$, a decision maker considers a state $\omega'$ possible if there is a choice problem for which state $\omega'$ ``makes a difference'' to the decision maker at state $\omega$. More precise, there are two acts that are identical except for the state $\omega'$ but the decision maker is not indifferent between those acts with her preferences in $\omega$.

Our aim is to analyze when $\tilde{\Pi}$ is a possibility correspondence satisfying properties 0. to 5. To this extent, we need to consider the following properties on $\succsim$.

We slightly abuse notation and denote by $S_{\omega}$ the space $S \in \mathcal{S}$ for which $\omega \in S$. Since any two spaces in $\mathcal{S}$ are disjoint, $S_{\omega}$ is unique, for all $\omega \in \Omega$.

\begin{ppty}\label{confined} If $f_{\{\omega'\}} g \succ(\omega) \ h_{\{\omega'\}} g$ for some $f, g, h \in \mathcal{A}$ and $S_{\omega''} \neq S_{\omega'}$, then $f'_{\{\omega''\}} g' \sim(\omega) \ h'_{\{\omega''\}} g'$ for all $f', g', h' \in \mathcal{A}$.
\end{ppty}

\begin{ppty}\label{nach_unten} If $f_{\{\omega'\}} g \succ(\omega) \ h_{\{\omega'\}} g$ for some $f, g, h \in \mathcal{A}$, then $S_{\omega'} \preceq S_{\omega}$.
\end{ppty}
These two properties characterize Confinement of the preference-based possibility correspondence.

\begin{lem}\label{confinementproof} For all $\omega \in \Omega$, $\succsim(\omega)$ satisfies Properties~\ref{confined} and~\ref{nach_unten} if and only if $\tilde{\Pi}$ satisfies Confinement.
\end{lem}

Close inspection of Confinement reveals that it consists of two properties of the possibility correspondence. First, for each state $\omega$ it confines the value of the correspondence to a single space. Second, it requires that this space is weakly lower than the space $S_{\omega}$. The proof of Lemma~\ref{confinementproof} in the appendix reveals that these two features are mimicked by Properties~\ref{confined} and~\ref{nach_unten}, respectively.

Property~\ref{confined} will play a role also in characterizing other properties of the possibility correspondence. For instance, whenever we write $\tilde{\Pi}^\uparrow(\omega)$, we implicitly assume that there is a well-defined base-space $S \supseteq \tilde{\Pi}(\omega)$.

Next we turn to Generalized Reflexivity.

\begin{ppty}\label{gen_reflexivity} $f_{\{\omega_S\}} g \succ(\omega) \ h_{\{\omega_S\}} g$ for some $S \preceq S_{\omega}$ and some $f, g, h \in \mathcal{A}$.
\end{ppty}

\begin{lem}\label{Gen_reflexivityproof} Suppose that Property~\ref{confined} holds. For all $\omega \in \Omega$, $\succsim(\omega)$ satisfies Property~\ref{gen_reflexivity} if and only if $\tilde{\Pi}$ satisfies Generalized Reflexivity.
\end{lem}

The proof is contained in the appendix.

Morris (1996, Lemma 3) showed that $f_{\{\omega\}} g \succ(\omega) \ h_{\{\omega\}} g$ for some $f, g, h \in \mathcal{A}$ and for all $\omega \in \Omega$ characterizes the truth axiom of knowledge in standard states-space structures. Property~\ref{gen_reflexivity} and Lemma~\ref{Gen_reflexivityproof} should be understood as generalizing this observation to unawareness structures.

\begin{ppty}\label{transitivity} If $f_{\{\omega'\}} g \succ(\omega) \ h_{\{\omega'\}} g$ for some $f, g, h \in \mathcal{A}$ and $f'_{\{\omega''\}} g' \succ(\omega') \ h'_{\{\omega''\}} g'$ for some $f', g', h' \in \mathcal{A}$, then $f''_{\{\omega''\}} g'' \succ(\omega) \ h''_{\{\omega''\}} g''$ for some $f'', g'', h'' \in \mathcal{A}$.
\end{ppty}

\begin{ppty}\label{euclideanness} If $f_{\{\omega'\}} g \succ(\omega) \ h_{\{\omega'\}} g$ for some $f, g, h \in \mathcal{A}$ and $f'_{\{\omega''\}} g' \succ(\omega) \ h'_{\{\omega''\}} g'$ for some $f', g', h' \in \mathcal{A}$, then $f''_{\{\omega''\}} g'' \succ(\omega') \ h''_{\{\omega''\}} g''$ for some $f'', g'', h'' \in \mathcal{A}$.
\end{ppty}

\begin{lem}\label{stationarityproof} For all $\omega \in \Omega$, $\succsim(\omega)$ satisfies Properties~\ref{transitivity} and~\ref{euclideanness} if and only if $\tilde{\Pi}$ satisfies Stationarity.
\end{lem}

Note that $\tilde{\Pi}$ satisfies stationarity if and only if it satisfies Transitivity (i.e., if $\omega' \in \tilde{\Pi}(\omega)$ then $\tilde{\Pi}(\omega') \subseteq \tilde{\Pi}(\omega)$) and Euclideanness (i.e., if $\omega' \in \tilde{\Pi}(\omega)$ then $\tilde{\Pi}(\omega') \supseteq \tilde{\Pi}(\omega)$). The proof contained in the appendix verifies that Property~\ref{transitivity} is a ``translation'' of Transitivity and that Property~\ref{euclideanness} is a ``translation'' of Euclideanness.

The next two properties relate preferences across spaces. E.g., they relate a decision maker's preference at $\omega$ to his preference at the projection to a lower space $\omega_S$. The proofs of the following two lemmata are contained in the appendix.

\begin{ppty}\label{PPI} If $\omega \in S'$, $S \preceq S'$, and $f_{\{\omega'\}} g \succ(\omega) \ h_{\{\omega' \}} g$ for some $f, g, h \in \mathcal{A}$, then there exists $\omega'' \in \Omega$ with $\omega' \in \{\omega''\}^{\uparrow}$ such that $f'_{\{\omega'' \}} g' \succ(\omega_S) \ h'_{\{\omega''\}} g'$ for some $f', g', h' \in \mathcal{A}$.
\end{ppty}

\begin{lem}\label{PPIproof} Suppose that Property~\ref{confined} holds. For all $\omega \in \Omega$, $\succsim(\omega)$ satisfies Property~\ref{PPI} if and only if $\tilde{\Pi}$ satisfies Projections Preserve Ignorance.
\end{lem}

\begin{ppty}\label{PPK} If $S \preceq S_{\omega'} \preceq S_{\omega}$ and $f_{\{\omega'\}} g \succ(\omega) \ h_{\{\omega'\}} g$ for some $f, g, h \in \mathcal{A}$, then $f'_{\{\omega''\}} g' \succ(\omega_S) \ h'_{\{\omega''\}} g'$ for some $f', g', h' \in \mathcal{A}$ if and only if there exists $\omega''' \in S_{\omega'}$ such that $\omega'' = \omega_S'''$ and $f''_{\{\omega'''\}} g'' \succ(\omega) \ h''_{\{\omega'''\}} g''$ for some $f'', g'', h'' \in \mathcal{A}$.
\end{ppty}

\begin{lem}\label{PPKproof} Suppose that Property~\ref{confined} holds. For all $\omega \in \Omega$, $\succsim(\omega)$ satisfies Property~\ref{PPK} if and only if $\tilde{\Pi}$ satisfies Projections Preserve Knowledge.
\end{lem}

We summarize our observations in the following proposition:

\begin{prop} The preference--based correspondence $\tilde{\Pi}$ is a possibility correspondence satisfying 0. to 5. if and only if for any $\omega \in \Omega$ the preference order $\succsim(\omega)$ satisfies Properties~\ref{confined} to~\ref{PPK}.
\end{prop}

\subsection{Preference-Based Knowledge and Awareness}

Assume now that for any $\omega \in \Omega$, $\succsim(\omega)$ is a preference relation satisfying Properties~\ref{confined} to~\ref{PPK}. We now define knowledge, awareness, and unawareness directly in terms of preferences.

\begin{defin}\label{Ktilde} The preference-based knowledge operator on $\Sigma$ is defined by
\begin{eqnarray*} \tilde{K}(E) := \left\{ \omega \in \Omega : \begin{array}{l} \mbox{ (i) } f_{\neg E} g \sim(\omega) \ h_{\neg E} g \mbox{ for all } f, g, h \in \mathcal{A}, and \\ \mbox{ (ii) } f_{E} g \succ(\omega) \ h_{E} g \mbox{ for some } f, g, h \in \mathcal{A}  \end{array}\right\}
\end{eqnarray*} if there is a state $\omega \in \Omega$ such that (i) and (ii) hold, and by $\tilde{K}(E) := \emptyset^{S(E)}$ otherwise.
\end{defin}

Property (i) is familiar from the definition of Savage null event (Savage, 1954). An event $E$ is known if its negation is null. Morris (1996, 1997) defines ``standard'' knowledge by this property alone.\footnote{To be precise, Morris (1996, 1997) defines ``Savage belief'' by property (i) alone. Since in our context we impose Property~\ref{gen_reflexivity} which implies the Truth, we think it is justified to call it knowledge.} Yet, in unawareness structures it is possible that both the event and its negation are null, in which case - as we will see below - the decision maker is unaware of the event. That's why we need to add the second requirement, property (ii), to the definition.

Given the possibility correspondence $\tilde{\Pi}$ and Definition~\ref{K}, the knowledge operator on $\Sigma$ is
\begin{eqnarray*} K(E) = \{ \omega \in \Omega : \tilde{\Pi}(\omega) \subseteq E \}
\end{eqnarray*} if there is a state $\omega \in \Omega$ such that $\tilde{\Pi}(\omega) \subseteq E$, and $K(E) = \emptyset^{S(E)}$ otherwise.

\begin{prop}\label{preference_based_K} For any event $E \in \Sigma$, $\tilde{K}(E) = K(E)$.
\end{prop}

The proof is contained in the appendix.\footnote{\label{finite}The proof makes use of the assumption that $\Omega$ is finite. We were not be able to prove it for more general $\Omega$ without imposing further assumptions. We suspect that this is one of the reasons why Morris (1996, 1997) assumed $\Omega$ to be finite.}

\begin{defin}\label{Atilde} The preference-based awareness operator on $\Sigma$ is defined by
\begin{eqnarray*} \tilde{A}(E) := \left\{ \omega \in \Omega : \begin{array}{l} \mbox{ (ii) } f_{E} g \succ(\omega) \ h_{E} g \mbox{ for some } f, g, h \in \mathcal{A}, or \\ \mbox{ (iii) } f_{\neg E} g \succ(\omega) \ h_{\neg E} g \mbox{ for some } f, g, h \in \mathcal{A}  \end{array}\right\}
\end{eqnarray*} if there is a state $\omega \in \Omega$ such that (ii) or (iii) hold, and by $\tilde{A}(E) := \emptyset^{S(E)}$ otherwise.
\end{defin}

Given the possibility correspondence $\tilde{\Pi}$ and Definition~\ref{A}, the decision maker's \emph{awareness operator} in $\Sigma$ is
\begin{equation*}
A(E) = \left \{\omega \in \Omega : \tilde{\Pi}(\omega) \subseteq
S(E)^{\uparrow} \right\}
\end{equation*}
if there is a state $\omega$ such that $\tilde{\Pi}(\omega) \subseteq
S(E)^{\uparrow}$, and $A(E) = \emptyset^{S(E)}$ otherwise.

\begin{prop} For any event $E \in \Sigma$, $\tilde{A}(E) = A(E)$.
\end{prop}

The proof follows from Proposition~\ref{preference_based_K} and Weak Necessitation.

\begin{defin}\label{Utilde} The preference-based unawareness operator on $\Sigma$ is defined by
\begin{eqnarray*} \tilde{U}(E) := \left\{ \omega \in \Omega : \begin{array}{l} \mbox{ (0) } f_{E} g \sim(\omega) \ h_{E} g \mbox{ for all } f, g, h \in \mathcal{A}, and \\ \mbox{ (i) } f_{\neg E} g \sim(\omega) \ h_{\neg E} g \mbox{ for all } f, g, h \in \mathcal{A}  \end{array}\right\}
\end{eqnarray*} if there is a state $\omega \in \Omega$ such that (0) and (i) hold, and by $\tilde{U}(E) := \emptyset^{S(E)}$ otherwise.
\end{defin}

This behavioral definition of unawareness means that a decision maker is unaware of an event if this event is null and the negation of the event is null.\footnote{This condition characterizes behaviorally unawareness in Schipper (2012) using a states-of-nature lattice structure and the Anscombe-Aumann approach.} Indeed, this behavioral definition of unawareness is equivalent to the epistemic notion of unawareness.

\begin{cor} For any event $E \in \Sigma$, $\tilde{U}(E) = U(E)$.
\end{cor}

Note that we are able to characterize all epistemic operators by conjunction and disjunction of choices experiments corresponding to properties (0) to (iii) in Definitions~\ref{Ktilde} to~\ref{Utilde}.

\section{Discussion}

In what sense do our results really provide a \emph{behavioral} characterization of properties of the possibility correspondence, knowledge, awareness, and unawareness? First note that in order to ``reveal'' whether a decision maker's possibility correspondence satisfies properties 0. to 5., many counterfactual choice experiments have to be performed. For instance, we first may have to conduct choice experiments at a state $\omega$ at which the decision maker may be unaware of both events $E$ and $F$. Then we may have to consider choice experiments at a state $\omega'$, at which the decision maker may be aware of $E$ but unaware of $F$, and then at yet another state $\omega''$ at which he is unaware of $E$ but aware of $F$ etc. But how should this be practically done? If these choice experiments are conducted sequentially, then it requires that after becoming aware of the $E$ the decision maker must become unaware of $E$ before we can conduct the last set of choice experiments.\footnote{Another difficulty that this model shares with Morris (1996, 1997) is how to phrase practically bets on events corresponding to (even higher order) beliefs of some events.}  More natural is the interpretation of the model as the \emph{belief of another player (or the modeler)} about the decision maker's choices at various states. While this interpretation may appear unusual at a first glance, we would argue that it corresponds rather closely to the actual treatment of decision theoretic models by decision theorists. Despite the emphasis on preferences revealed by choices, most decision theorists chose not to conduct choice experiments and many choice theoretic ``axioms'' are just of technical nature (e.g., continuity axioms) or impractical to implement in actual choice experiments (e.g., axioms on conditional preferences). Such ``axioms'' are testable just in ``principle''. That is, they represent the belief of the modeler about how a decision maker would perform in certain choice experiments when he is ascribed a certain utility representation.\footnote{Our interpretation of the decision theoretic model is also reminiscent of the interpretation of the notion of strategy in extensive-form games. As Rubinstein (1991) pointed out, an action prescribed by a strategy at an information set which is excluded by an earlier move of that very strategy is implicitly interpreted in game theory as the beliefs that the other players entertain regarding the player's move if that information set were reached.}

Our analysis is subject to another caveat raised previously in Schipper (2012). Property 1 implies that the events of which the decision maker is unaware of do not affect her ranking of acts. This holds even for composite acts conditioned on events that the decision maker is unaware of. More generally, it rules out that a decision maker becomes aware of an event merely from facing an act. This is also the implicit assumption in standard decision theory (i.e., different acts do not change the subset of ``small worlds'' that a decision maker pays attention to). Yet, there it is less problematic since there is just one set of ``small worlds''. But it may be unrealistic in some situations with unawareness. Sometimes, when facing an act, a decision maker may become in very subtle ways a bit more careful with the ``fine prints'' of acts, and this care may lead her to become aware of events. E.g., a buyer facing a decision about whether or not to enter into a certain purchase contract may become aware of events that she was previously unaware of when reading all the fine prints of the contract. If ex ante an outside observer does not know how acts affect the awareness of a decision maker, then it may be impossible to design choice experiments required to elicit properties of the static possibility correspondence.

Our analysis differs from the one presented in Schipper (2012) in several aspects. First, we mentioned already that Schipper (2012) considers states of nature while we consider states of the world. Latter approach allows us to talk about knowledge of knowledge etc. Second, while Schipper (2012) considers Anscombe-Aumann acts, i.e., mappings from states to mixtures over outcomes, we consider here real-valued Savage acts, i.e., mappings from states to real numbers. Third, Schipper (2012) aims at subjective expected utility representation theorems with probabilistic beliefs while we focus on characterizing the qualitative notions of knowledge, awareness, and unawareness. Our analysis is related to Li (2008) who seeks to study the difference between unawareness and zero probability as well.

Morris (1996) also related dynamic preferences to properties of belief change. Dynamic unawareness has been analyzed in epistemic frameworks (van Dittmarsch and French 2009, 2011a, b, Hill, 2010), in game theoretic settings (Feinberg, 2012, Grant and Quiggin, 2013, Halpern and R\^{e}go, 2012b, R\^{e}go and Halpern, 2012, Heifetz, Meier, and Schipper, 2011a, b, Li, 2006, Meier and Schipper, 2011), and more recently also in a decision theoretic framework by Karni and Vier{\o} (2013). The latter approach is confined to states of nature, i.e., the beliefs of the decision maker are not part of the description of the state. We leave a decision theoretic analysis of dynamic unawareness with states of the world analogous to Morris (1996) to future research.

\appendix

\section{Proofs}

\subsection{Proof of Lemma~\ref{confinementproof}}

``$\Rightarrow$'': Suppose there exists $S \in \mathcal{S}$ and $\omega \in S$ such that $\tilde{\Pi}(\omega) \subseteq S'$ with $S' \npreceq S$. Then for $\omega' \in \tilde{\Pi}(\omega)$ we have $f_{\{\omega'\}} g \succ(\omega) \ h_{\{\omega'\}} g$ for some $f, g, h \in \mathcal{A}$. But Property~\ref{nach_unten} implies $S_{\omega'} \preceq S_{\omega}$. Note that $S' = S_{\omega'}$, a contradiction.

Suppose now that there exists $S \in \mathcal{S}$ and $\omega \in S$ such that $\tilde{\Pi}(\omega) \nsubseteq S'$, for all $S' \in \mathcal{S}$. Then there exist $\omega', \omega'' \in \tilde{\Pi}(\omega)$ such that $S_{\omega'} \neq S_{\omega''}$. If $\omega' \in \tilde{\Pi}(\omega)$ then by Property~\ref{confined} we have $f_{\{\omega'\}} g \succ(\omega) \ h_{\{\omega'\}} g$ for some $f, g, h \in \mathcal{A}$ and for any $\omega'' \neq \omega'$ with $S_{\omega''} \neq S_{\omega'}$, $f'_{\{\omega''\}} g' \sim(\omega) \ h'_{\{\omega''\}} g'$ for all $f', g', h' \in \mathcal{A}$. Hence $\omega'' \notin \tilde{\Pi}(\omega)$, a contradiction.

``$\Leftarrow$'': Suppose there exist $\omega', \omega'' \in \Omega$ such that Property~\ref{confined} is violated, i.e., $f_{\{\omega'\}} g \succ(\omega) \ h_{\{\omega'\}} g$ for some $f, g, h \in \mathcal{A}$, $S_{\omega''} \neq S_{\omega'}$, and $f'_{\{\omega''\}} g' \nsim(\omega) \ h'_{\{\omega''\}} g'$ for some $f', g', h' \in \mathcal{A}$. Then $\omega', \omega'' \in \tilde{\Pi}(\omega)$, a contradiction to $\tilde{\Pi}(\omega) \subseteq S'$ for some $S' \in \mathcal{S}$.

Suppose there exist $\omega, \omega', \omega'' \in \Omega$ such that Property~\ref{nach_unten} is violated, i.e., $f_{\{\omega'\}} g \succ(\omega) h_{\{\omega'\}} g$ for some $f, g, h \in \mathcal{A}$ and $S_{\omega'} \npreceq S_{\omega}$. Then by previous arguments $\tilde{\Pi}(\omega) \subseteq S'$ with $S' = S_{\omega'} \npreceq S_{\omega} = S$, a contradiction of Confinement.\hfill $\Box$

\subsection{Proof of Lemma~\ref{Gen_reflexivityproof}}

``$\Leftarrow$'': Suppose by contradiction that $f_{\{\omega_S\}} g \sim(\omega) \ h_{\{\omega_S\}}$ for all $S \preceq S_{\omega}$ and all $f, g, h \in \mathcal{A}$. Thus, $r^{S_{\omega}}_{S}(\omega) \notin \tilde{\Pi}(\omega)$ for all $S \preceq S_{\omega}$, a contradiction.

``$\Rightarrow$'': Suppose $f_{\{\omega_S\}} g \succ(\omega) \ h_{\{\omega_S\}} g$ for some $S \preceq S_{\omega}$ and some $f, g, h \in \mathcal{A}$. Thus, $\omega_S \in \tilde{\Pi}(\omega)$. Property~\ref{confined} implies $\tilde{\Pi}(\omega) \subseteq S$. Thus, $\Pi^{\uparrow}(\omega)$ is well-defined and we conclude that $\omega \in \tilde{\Pi}^{\uparrow}(\omega)$.\hfill $\Box$

\subsection{Proof of Lemma~\ref{stationarityproof}}

We first prove that Transitivity is equivalent to Property~\ref{transitivity}: $f_{\{\omega''\}} g \succ(\omega) \ h_{\{\omega''\}} g$ for some $f, g, h \in \mathcal{A}$ if and only if $\omega'' \in \tilde{\Pi}(\omega)$. $f'_{\{\omega'\}} g' \succ(\omega) \ h'_{\{\omega'\}} g'$ for some $f', g', h' \in \mathcal{A}$ if and only if $\omega' \in \tilde{\Pi}(\omega)$. $f''_{\{\omega''\}} g'' \succ(\omega') \ h''_{\{\omega''\}} g''$ for some $f'', g'', h'' \in \mathcal{A}$ if and only if $\omega'' \in \tilde{\Pi}(\omega')$. Thus, Property~\ref{transitivity} is equivalent to if $\omega' \in \tilde{\Pi}(\omega)$ and $\omega'' \in \tilde{\Pi}(\omega')$, then $\omega'' \in \tilde{\Pi}(\omega)$. This is equivalent to Transitivity.

Second, we prove that Euclideanness is equivalent to Property~\ref{euclideanness}: $f_{\{\omega''\}} g \succ(\omega') \ h_{\{\omega''\}} g$ for some $f, g, h \in \mathcal{A}$ if and only if $\omega'' \in \tilde{\Pi}(\omega')$. $f'_{\{\omega'\}} g' \succ(\omega) \ h'_{\{\omega'\}} g'$ for some $f', g', h' \in \mathcal{A}$ if and only if $\omega' \in \tilde{\Pi}(\omega)$. $f''_{\{\omega''\}} g'' \succ(\omega) \ h''_{\{\omega''\}} g''$ for some $f'', g'', h'' \in \mathcal{A}$ if and only if $\omega'' \in \tilde{\Pi}(\omega)$. Thus, Property~\ref{euclideanness} is equivalent to if $\omega' \in \tilde{\Pi}(\omega)$ and $\omega'' \in \tilde{\Pi}(\omega)$, then $\omega'' \in \tilde{\Pi}(\omega')$. This is equivalent to Euclideanness.\hfill $\Box$

\subsection{Proof of Lemma~\ref{PPIproof}}

``$\Rightarrow$'': If $f_{\{\omega'\}} g \succ(\omega) \ h_{\{\omega'\}} g$ for some $f, g, h \in \mathcal{A}$, then $\omega' \in \tilde{\Pi}(\omega)$. If $f'_{\{\omega''\}} g' \succ(\omega_S) \ h'_{\{\omega''\}} g'$ for some $f', g', h' \in \mathcal{A}$, then $\omega'' \in \tilde{\Pi}(\omega_S)$. Moreover, according to Property~\ref{PPI}, $\omega' \in \{\omega''\}^{\uparrow}$. From Property~\ref{confined} follows that $\tilde{\Pi}^{\uparrow}(\omega)$ and $\tilde{\Pi}^{\uparrow}(\omega_S)$ are well-defined. Thus $\omega' \in \tilde{\Pi}^{\uparrow}(\omega_S)$ and it follows that $\tilde{\Pi}^{\uparrow}(\omega) \subseteq \tilde{\Pi}^{\uparrow}(\omega_S)$.

``$\Leftarrow$'': $\omega' \in \tilde{\Pi}(\omega)$ if and only if $f_{\{\omega'\}} g \succ(\omega) \ h_{\{\omega'\}} g$ for some $f, g, h \in \mathcal{A}$. Since $\omega' \in \tilde{\Pi}^{\uparrow}(\omega_S)$,  by Projections Preserve Ignorance there exists $\omega'' \in \tilde{\Pi}(\omega_S)$ such that $\omega' \in \{\omega''\}^{\uparrow}$. Moreover, $f'_{\{\omega''\}} g' \succ(\omega_S) \ h'_{\{\omega''\}} g'$ for some $f', g', h' \in \mathcal{A}$\hfill $\Box$

\subsection{Proof of Lemma~\ref{PPKproof}}

Let $S \preceq S_{\omega'} \preceq S_{\omega}$. By Property~\ref{confined}, $f_{\{\omega'\}} g \succ(\omega) \ h_{\{\omega'\}} g$ for some $f, g, h \in \mathcal{A}$ if and only if $\tilde{\Pi}(\omega) \subseteq S_{\omega'}$. Thus, the hypotheses of both Projections Preserve Knowledge and Property~\ref{PPK} are equivalent under Property~\ref{confined}.

$f'_{\{\omega''\}} g' \succ(\omega_S) \ h'_{\{\omega''\}} g'$ for some $f', g', h' \in \mathcal{A}$ if and only if $\omega'' \in \tilde{\Pi}(\omega_S)$.

$f''_{\{\omega'''\}} g'' \succ(\omega) \ h''_{\{\omega'''\}} g''$ for some $f'', g'', h'' \in \mathcal{A}$ if and only if $\omega''' \in \tilde{\Pi}(\omega)$.

Thus, the conclusion of Property~\ref{PPK} is equivalent to: $\omega'' \in \tilde{\Pi}(\omega_S)$ if and only there exists a state $\omega''' \in S_{\omega'}$ such that $\omega'' = \omega_S'''$ and $\omega''' \in \tilde{\Pi}(\omega)$. This is equivalent to $\tilde{\Pi}(\omega_S) = (\tilde{\Pi}(\omega))_S$, the conclusion of Projections Preserve Knowledge. \hfill $\Box$

\subsection{Proof of Proposition~\ref{preference_based_K}}

``$\supseteq$'': $\omega \in K(E)$ if and only if $\tilde{\Pi}(\omega) \subseteq E$. It follows that $f_{\{\omega'\}} g \succ(\omega) \ h_{\{\omega'\}} g$ for some $f, g, h \in \mathcal{A}$ implies $\omega' \in E$. Define
\begin{eqnarray*} f'(\omega) & = & \left\{ \begin{array}{cl} f(\omega) & \mbox{ if } \omega = \omega' \\ g(\omega) & \mbox{ if } \omega \neq \omega' \end{array} \right. \\ h'(\omega) & = & \left\{ \begin{array}{cl} h(\omega) & \mbox{ if } \omega = \omega' \\ g(\omega) & \mbox{ if } \omega \neq \omega' \end{array} \right.
\end{eqnarray*} and $g' = g$. Note that $f', h', g' \in \mathcal{A}$. Then from $f_{\{\omega'\}} g \succ(\omega) \ h_{\{\omega'\}} g$ for some $f, g, h \in \mathcal{A}$ implies $\omega' \in E$ follows that $f'_E g' \succ(\omega) \ h'_E g'$ for some $f', g', h' \in \mathcal{A}$, yielding property (ii).

Note that property (i) in the definition of $\tilde{K}$ holds trivially if $E = S(E)^{\uparrow}$. Assume $E \subsetneqq S(E)^{\uparrow}$. From $\tilde{\Pi}(\omega) \subseteq E$ follows that $\omega' \in \neg E$ implies $\omega' \notin \tilde{\Pi}(\omega)$. Thus $f_{\{\omega'\}} g \sim(\omega) \ h_{\{\omega'\}} g$ for all $f, g, h \in \mathcal{A}$. Since $\Omega$ is finite, we can enumerate all states in $\neg E$ from lets say $1$ to $n$. Then for any $f, g, h$ we have $f_{\{\omega_1\}} g \sim(\omega) \ f_{\{\omega_1, \omega_2\}} g \sim(\omega) \ ... \sim(\omega) \ f_{\{\omega_1, \omega_2, ..., \omega_n\}} g = f_{\neg E} g$ as well as $h_{\{\omega_1\}} g \sim(\omega) \ h_{\{\omega_1, \omega_2\}} g \sim(\omega) \ ... \sim(\omega) \ h_{\{\omega_1, \omega_2, ..., \omega_n\}} g = h_{\neg E} g$ and thus by transitivity of $\succeq(\omega)$ also $f_{\neg E} g \sim(\omega) \ h_{\neg E} g$, yielding property (i) of the definition of $\tilde{K}$. Hence $\omega \in \tilde{K}(E)$.

``$\subseteq$'': $\omega \in \tilde{K}(E)$ if and only if (i) $f_{\neg E} g \sim(\omega) \ h_{\neg E} g$ for all $f, g, h \in \mathcal{A}$, and (ii) $f_{E} g \succ(\omega) \ h_{E} g$ for some $f, g, h \in \mathcal{A}$. Consider first (i). For any $\omega' \in \neg E$, arguments made above imply $f_{\{\omega'\}} g \sim(\omega) \ h_{\{\omega'\}} g$ for all $f, g, h \in \mathcal{A}$ if we let $\omega' = \omega_1$. It follows that $\tilde{\Pi}(\omega) \cap \neg E = \emptyset$.

Consider now property (ii) of the definition of $\tilde{K}$. We claim that for some $\omega' \in E$, $f_{\{\omega'\}} g \succ(\omega) \ h_{\{\omega'\}} g$ for some $f, g, h \in \mathcal{A}$. Suppose to the contrary that $f_{\{\omega'\}} g \sim(\omega) \ h_{\{\omega'\}} g$ for all $f, g, h \in \mathcal{A}$ and all $\omega' \in E$. Since $\Omega$ is finite, we can enumerate states in $E$ from lets say $1$ to $m$. Previous arguments imply $f_E g \sim(\omega) \ h_E g$ for all $f, g, h \in \mathcal{A}$, a contradiction to property (ii). Thus $\tilde{\Pi}(\omega) \subseteq E$ which is equivalent to $\omega \in K(E)$. \hfill $\Box$


\end{document}